\documentclass[article,amsmath,fleqn,amsart,floats,nofootinbib,aps,twocolumn,pra,floatfix,showpacs,superscriptaddress]{revtex4-2}
\usepackage{cmap}
\usepackage{wrapfig} % Плавающие картинки
\usepackage{graphicx}
\usepackage{pgfplotstable}
\usepackage[pdftex,unicode,colorlinks,bookmarks=false,citecolor=blue,linkcolor=blue,urlcolor=blue]{hyperref} % нумерование страниц!!!! ИМЕННО В ТАКОМ ПОРЯДКЕ СО СЛЕДУЮЩИМ ПАКЕТОМ
\usepackage[T1,T2A]{fontenc} % Пакет выбора кодировки и шрифтов
\usepackage{latexsym}
\usepackage{amsmath}
\usepackage{amssymb}
\usepackage{amsfonts}
\usepackage{color}
\usepackage{bm}
\usepackage{verbatim}
\usepackage[english]{babel}
\usepackage[utf8]{inputenc}
\usepackage{euscript}
\usepackage{subfigure}
\usepackage[normalem]{ulem}
\usepackage{listings}
\hypersetup{
colorlinks=true,
linkcolor=blue,
filecolor=magenta,
urlcolor=cyan,
}

\begin{document}
\title{Nonequilibrium effects in ballistic point contacts $Ta-Cu$ and $2H-NbS{{e}_{2}}-Cu$. Two-gap superconductivity in $2H-NbS{{e}_{2}}$}

\author{N.L.~Bobrov}
\affiliation{B.I.~Verkin Institute for Low Temperature Physics and Engineering of the National Academy of Sciences of Ukraine\\ 47 Nauka ave., 61103 Kharkov, Ukraine}

\email{bobrov@ilt.kharkov.ua}

\published {\href{https://doi.org/10.1007/s10909-024-03052-x}{J Low Temp Phys \textbf{214} 399–426 (2024).}}

\begin{abstract}{The heterocontacts $Ta-Cu$ and $NbS{{e}_{2}}-Cu$ have been studied. For the $Ta-Cu$ contacts the theoretical estimation of the value of $\delta$-functional barrier at the boundary arising due to mismatch of fermionic parameters of the contacting metals is carried out and a good agreement between the calculation and experiment is obtained. An expression for the estimation of the diameter of the heterocontact on either side of the boundary is obtained. The magnitude of the jump-like decrease in the excess current (and the superconducting gap) due to the phase transition of the superconductor region near the point contact into a spatially inhomogeneous state when the critical concentration of nonequilibrium quasiparticles is reached has been determined. Obtained dependence of the additive differential resistance on the displacement at the contact arising after the phase transition, due to the excess charge of quasiparticles and the associated reverse current (or additive voltage). In $2H-NbS{{e}_{2}}$ there is a two-zone superconductivity character with $\sim8$ times different energy gap values. Under the influence of current injection of nonequilibrium quasiparticles there is a sequential phase transition of layers adjacent to the point contact, in a spatially inhomogeneous state with a suppressed gap, which is accompanied by a step change in the slope of the $I-V$ curve with a discrete increase in differential resistance, jump-like movement of the boundary between areas with suppressed and equilibrium values of the energy gaps.\\

 }.

\end{abstract}

\maketitle

\tableofcontents{}
\newpage

\section{Introduction}\label{sec1}

The emergence of spatially inhomogeneous states in three-dimensional superconducting ballistic point contacts is one of the most interesting and understudied problems in the field of nonequilibrium superconductivity. In $S-c-N$ point contacts, in many cases there is no local equilibrium between electrons and phonons in the current concentration region. There is also no equilibrium between quasiparticle excitations and condensate in the superconducting bank. The $I-V$ curves  of the $S-c-N$ point contact is significantly nonlinear.

In addition to the nonlinearity at $eV\sim\Delta$ displacements due to the energy gap, in some cases there may be a comparable intensity nonlinearity of non-spectral character due to nonequilibrium processes in the near-contact region.

So far, two \cite{1,2} papers have been published on this topic. They present the results of the study of the evolution of nonequilibrium features in the spectra of tantalum-based contacts -- temperature and magnetofield dependences, as well as resistance dependences.

The theory for ballistic $S-c-N$ point contacts \cite{3} predicts that the second derivatives of the $I-V$ curves of such contacts at the transition from the normal to the superconducting state at $eV \gg \Delta$ change very slightly. The additional nonlinearity arising at the transition is due to inelastic processes involving the emission of phonons during the Andreev reflection of electrons, which form an excess current (defined here as the difference between the $I-V$ curves in the $S$- and $N$-states). Such processes are sufficiently effective only in the region of high current density, where the concentration of interacting quasiparticles is sufficiently large. It is much smaller than the nonlinearity of the normal state due to backscattering processes (backscattering processes are such scattering acts in which the electron, after an inelastic collision with a phonon, returns to the same half-space from which it flew out).

It follows from the \cite{3} theory that after the transition of the contact to the superconducting state, the phonon features in the spectrum shift to lower energies by an order of magnitude of $\Delta$, and their amplitude slightly decreases due to additional broadening by the same magnitude. Experiments on tin point contacts brilliantly confirmed the conclusions of the theory. In Fig. 3 of the \cite{4} shows the second derivatives for the $Sn-Cu$ heterocontact in the superconducting and normal states, where the metamorphosis predicted by the theory can be observed.

Tantalum, as well as tin, is a superconductor of the first kind with sufficiently close values of the superconducting energy gap and critical temperature. However, the reduced coherence length $\zeta$ (the coherence length taking into account the elastic relaxation length of electrons) in these metals is very different; the corresponding estimate is given in \cite{5}. It follows from this estimate that the volume of the sphere, limited by the reduced coherence length, in tantalum is smaller than in tin $\sim100$ times. Since the minimum volume of the superconductor in which the energy gap can vary cannot be smaller than this value, we can expect that not only the region of high current density near the hole, but also the region of the order of $\zeta$, which is not taken into account by the \cite{3} theory, will participate in the formation of the nonlinearity of the $I-V$ curves in the superconducting state, which was confirmed in the experiments.

For tantalum-based ballistic point contacts, the spectrum undergoes radical changes during the transition to the superconducting state. Instead of broadening, the phonon peaks sharply sharpen and their amplitude increases. Especially strong changes occur in the low-frequency region of the phonon spectrum. Fig.7(a) shows in \cite{2} the successive evolution of the second derivative of the $I-V$ curve of the point contact at decreasing temperature. Note that in addition to the gradual sharpening of phonon peaks, a feature (indicated by the arrow) appears in place of the soft phonon mode near 7 mV on the curve 2 of \cite{2}, whose intensity increases rapidly with decreasing temperature, and its position shifts to the region of lower energies.

We briefly discuss the reason for the sharpening of the phonon peaks and then return to the discussion of the new feature. The contribution to the phonon features associated with the excess current is a superposition of contributions from two spatially separated regions: the contribution from the region with high current density, which corresponds to the theoretical model, and the contribution from the region with a size on the order of the coherence length, which is responsible for the sharpening of the phonon peaks.

In principle, the mechanism of influence on the phonon features from both regions is the same - it is the reabsorption of nonequilibrium phonons by Andreev electrons. But since the second region is much larger than the contact diameter, the difference in the group velocities of nonequilibrium phonons generated by the electrons begins to play a significant role. The minimum group velocity is observed near the maxima of the density of phonon states. These phonons accumulate in the indicated volume, which provides a higher probability of interaction with Andreev electrons and Cooper pairs, providing the largest addition to the spectrum. For phonon velocity selection, fast phonons must be free to leave this volume; the phonon passage mode must be ballistic. The smaller Fermi velocity of carriers in tantalum compared to tin also plays an important role.

The position of the new feature on the voltage axis depends not only on the temperature, but also on the resistance and the value of the magnetic field. At a fixed temperature, the voltage at which the feature is observed is proportional to $R^{1/2}$ (see Fig. 8 in \cite{2}), and corresponds to the constancy of the critical power when the contact resistance changes by almost an order of magnitude \cite{2} (at $T=2K$ $P_c \sim 0.4 \mu W$). The shape of the feature corresponds to a stepwise decreasing excess current.

Immediately after the singularity, the differential resistance of the contact increases and becomes greater than the differential resistance in the normal state at the same voltage (see Fig.\hyperref[Fig3]{3} below). Fig.10(a) shows in \cite{2} the dependence of the intensity and position of the feature on the energy axis on the magnetic field. It can be seen that as the magnetic field increases, the amplitude first decreases sharply and then more smoothly and shifts to the region of higher energies. The critical power increases with temperature or value of the magnetic field, which excludes the interpretation of the singularity as a result of destruction of superconductivity by temperature or magnetic field and suggests a different character of its origin.

To interpret such unusual dependences in the \cite{1,2}, a phenomenological model was proposed to explain this feature as a transition of the superconductor region adjacent to the point contact into a nonequilibrium state with a suppressed gap when a critical concentration of excess quasiparticles is reached, similar to what occurs in experiments with nonequilibrium quasiparticle injection in \cite{6} tunnel structures.

The current flowing through the contact is a source of nonequilibrium quasiparticles, which are able to multiply by reabsorption of nonequilibrium phonons and electron-electron collisions in such a way that most of the quasiparticles acquire excess energy $\sim\Delta$, accumulating in the layer above the ceiling of the energy gap. Taking into account that the final excess energy of quasiparticles is of the order of $\Delta$, at the same power per unit time approximately equal number of excess quasiparticles above the slit will be generated, which explains the constancy of the critical power of the non-equilibrium feature. 

Now let us consider the reasons for the increase of the critical injection power with increasing temperature or magnetic field. The stationary density of excess quasiparticles depends on their relaxation rate increasing with temperature and field. To compensate this increase, the critical injection current $I_c=V_c/R$ must increase.

Let us give a qualitative explanation from the \cite{2} of the transformation of PC spectra at the transition of one of the electrodes into the superconducting state. Quasiparticles with maximum energy $eV\approx\hbar\omega_D\gg\Delta$ are injected into the superconductor through the hetero-boundary. These quasiparticles populate the electron- or hole-like branch of the excitation spectrum, depending on the polarity of the applied voltage. The excitations relax relatively quickly, emitting phonons, and accumulate in a layer of the order of $\Delta$ above the ceiling of the energy gap.

Further relaxation of the residual population imbalance of the excess quasiparticles occurs rather slowly, over a time $\tau_0\sim\tau_{ep}(\Delta)$ \cite{7}. During this time, the quasi-particles diffuse deep into the conductor, creating excess charge and an associated reverse current or added voltage that increases the contact resistance. Factors that reduce the value of the unbalance (inelastic scattering by phonons, superconducting current, or an external magnetic field in the presence of scattering by impurities) reduce the reverse current and contact resistance. For all spectra, the effect should decrease with increasing bias $eV$, which is what is observed in the experiment.

To summarize: the current flowing through the contact is a source of nonequilibrium quasiparticles. These quasiparticles are able to multiply by emitting phonons absorbed by Cooper pairs. The stationary concentration $n$ of excitations in the neighborhood of the superconductor near the contact is determined by the ratio of the rates of generation and recombination of quasiparticles. The value of $n$ depends on the value of $\Delta$ in this neighborhood and the excess current proportional to the gap. When the critical injection power is reached, there is a jump-like transition of the superconductor region into a nonequilibrium state. The energy gap and the associated excess current decrease, the differential resistance after the nonequilibrium feature increases and becomes larger than the differential resistance of the point contact in the normal state at the same voltage, due to the reverse current increasing the contact resistance. With increasing voltage, the residual unbalance of the population imbalance of the excitation spectrum branch of excess quasiparticles decreases, which leads to a decrease in the differential resistance of the microcontact relative to the differential resistance in the normal state (see Fig. \hyperref[Fig3]{3}). 

We emphasize that the jump-like transition to a nonequilibrium state with a suppressed gap was observed only for an unperturbed superconductor with a perfect lattice. In dirty tantalum-based point contacts, in which phonon features are strongly blurred in the normal state, such features were not observed. The minimum volume of a superconductor capable of transitioning to a nonequilibrium state with a suppressed gap cannot be smaller than the coherence length. Therefore, for superconductors with a large coherence length, in the region of phonon energies the critical concentration is not reached and nonequilibrium features in the spectra are present.

\section{Theory}\label{sec2}

When analyzing experimental data in point-contact spectroscopy, one of the most important values is information about the contact size. Without knowledge of this parameter, it is often impossible to perform an objective analysis of physical processes occurring both in the current concentration region and in the peripheral regions of the contact, comparing the size with characteristic relaxation lengths of quasiparticles, coherence length, etc. For ballistic homocontacts, i.e., for contacts between identical metals, this problem has long been solved and the corresponding analytical expressions are available. The most widely used Sharvin's equation \cite{8} for the model of a contact in the form of a circular hole in an infinitely thin opaque flat wall separating two metallic half-spaces:

\label{eq1}
\begin{equation}
d=({16\rho l}/{3\pi R_0})^{1/2}.
\end{equation}

Here $\rho l$ - the product of specific electrical resistance by the mean free path length of electrons is a constant characteristic of each particular metal. In the free-electron approximation
\begin{equation}
\label{eq2}
\rho l={{p}_{F}}/n{{e}^{2}}=\frac{3{{\pi }^{2}}\hbar }{k_{F}^{2}{{e}^{2}}};
\end{equation} 
then
\label{eq3}
\begin{equation}
d=\frac{4}{e{{k}_{F}}}{{\left( \frac{\pi \hbar }{{{R}_{0}}} \right)}^{{1}/{2}\;}}.
\end{equation}

In basic theoretical work on point-contact spectroscopy, a relation for the contact resistance at zero bias (formula (24) in \cite{9}) is given, which corresponds to Sharvin's formula:

\begin{equation}
\label{eq4}
R_{0}^{-1}=\frac{{{e}^{2}}S{{S}_{F}}}{{{\left( 2\pi \hbar  \right)}^{3}}}{{\left\langle \alpha \right\rangle }_{\begin{smallmatrix} \\ {{v}_{z}}>0
\end{smallmatrix}}}.
\end{equation}
here $S$ is the area of an arbitrarily shaped hole, $S_F$ is the area of the Fermi surface (FS); $\left\langle ... \right\rangle {{v}_{z}}>0$ means averaging over half of the Fermi surface corresponding to $v_z>0$, $\alpha ={{{v}_{z}}}/{{{v}_{F}}=\cos \theta }\;$;
For a spherical FS $\left\langle \alpha \right\rangle {{v}_{z}}>0=1/2$, $S_F=4\pi p_{F}^2$, whence $R_0=4p_F/3e^{2}nS$ or in the circular hole model:
\begin{equation}
\label{eq5}
R_{0}=\frac{16\pi \hbar }{{{e}^{2}}k_{F}^{2}{{d}^{2}}}.
\end{equation}

Thus for ballistic homocontacts in the hole model, the contact resistance is determined by its geometry (area) and is called the constriction resistance.

Since superconductor-normal metal point contacts (hereafter $S-c-N$, here $c$ is a constriction) are always heterocontacts, let us first consider the general regularities inherent to them in the normal state and then proceed to how they will manifest themselves in the transition to the superconducting state. We will consider only ballistic point contacts that do not have any additional scatterers in the form of impurities, defects, etc. at the boundary. Most of the equations below are taken from an earlier publication \cite{10} devoted to finding the point-contact electron-phonon interaction function ("EPI") in tantalum, including the use of heterocontacts.

As follows from the theory ( Equation 10 in \cite{11}), at direct contact between metals, the resistance of the heterocontact is determined not only by the constriction resistance, but also by the barrier at the hetero-boundary, arising due to the difference in the Fermi parameters of the contacting metals.  

The resistance of the heterocontact in the presence of such a barrier at the hetero-boundary is the same when averaged over any of the metals forming the heterocontact, which follows from the continuity of the $z$-component of the current density vector and does not depend on the dispersion law.

\begin{equation}
\label{eq6}
R_{het}^{-1}=\frac{{{e}^{2}}S{{S}_{F}}}{{{\left( 2\pi \hbar  \right)}^{3}}}{{\left\langle \alpha D(\alpha ) \right\rangle }_{\begin{smallmatrix} 1,2 \\ {{v}_{z}}>0
\end{smallmatrix}}}.
\end{equation}

As can be seen, the equations \hyperref[eq4]{4} and \hyperref[eq6]{6} are very similar and their corresponding notations are the same.
Here $\left\langle ... \right\rangle {{{v}_{z}}}>0$ means averaging over the Fermi surface of metal 1 or 2, respectively, under the condition ${{v}_{z}}>0$; $S$ -- contact area; ${{S}_{F1,2}}$ the Fermi surface area of metal 1 or 2; $\alpha ={{{v}_{z}}}/{{{v}_{F}}=\cos \theta }\;$; $D$ -- boundary passing coefficient. Since $R_{het}^{-1}$ is independent of the metal number over which the averaging is performed, we can write:

\begin{equation}
\label{eq7}
{{\left\{ {{S}_{F}}\left\langle \alpha D(\alpha ) \right\rangle  \right\}}_{1}}={{\left\{ {{S}_{F}}\left\langle \alpha D(\alpha ) \right\rangle  \right\}}_{2}}.
\end{equation}

Combining equations \hyperref[eq4]{4} and \hyperref[eq6]{6} we obtain:

\begin{equation}
\label{eq8}
{{R}_{het}}={{R}_{0}}\left( {{{\left\langle \alpha  \right\rangle }_{{{v}_{z}}>0}}}/{{{\left\langle \alpha D(\alpha ) \right\rangle }_{{{v}_{z}}>0}}}\; \right).
\end{equation}

In a metal with a large value of $p_{F}$, the relative phase volume of nonequilibrium filled states is smaller due to the reflection of part of the electron trajectories from the hetero-boundary. The electron passage coefficient through the hetero-boundary for a spherical Fermi surface (since such quantities as the Fermi velocity cannot be clearly defined if the Fermi surface is not spherical \cite{12}) is:
\begin{equation}
\label{eq9}
D=\frac{4{{v}_{z1}}{{v}_{z2}}}{{{\left( {{v}_{z1}}+{{v}_{z2}} \right)}^{2}}},
\end{equation}
here ${{v}_{z1}}={{v}_{F1}}\cos {{\theta }_{1}}$; ${{v}_{z2}}={{v}_{F2}}\cos {{\theta }_{2}}$; $v_{F1,2}$ is the velocity, Fermi of metal 1 or 2. In the case where the angle of incidence of the electron on the hetero-boundary deviates from the vertical, the electron trajectory experiences refraction when crossing it. This is due to the fact that when the electron passes from one metal to another, the law of conservation of the tangential component of its momentum must be satisfied:
 $p_\parallel ={{p}_{F1}}\sin {{\theta }_{1}}={{p}_{F2}}\sin {{\theta }_{2}}$. Denote ${{{p}_{F1}}}/{{{p}_{F2}}=b}\;$; ${{{v}_{F1}}}/{{{v}_{F2}}=c}\;$; $\cos {{\theta }_{1}}={{\alpha }_{1}}$; $\cos {{\theta }_{2}}={{\alpha }_{2}}$. We assume for definiteness that $b<1$. As a result we obtain
\begin{equation}
\label{eq10}
{{\alpha }_{1}}={{b}^{-1}}{{\left( \alpha _{2}^{2}+{{b}^{2}}-1 \right)}^{{1}/{2}\;}};\ {{\alpha }_{2}}=b{{\left( \alpha _{1}^{2}+{{b}^{-2}}-1 \right)}^{{1}/{2}\;}}.
\end{equation}
The transmission coefficients at each bank can be written in the form:
\begin{equation}
\label{eq11}
D\left( {{\alpha }_{1}} \right)=\frac{4b{{\alpha }_{1}}{{\left( \alpha _{1}^{2}+{{b}^{-2}}-1 \right)}^{{1}/{2}\;}}}{c{{\left[ {{\alpha }_{1}}+\left( b/c \right){{\left( \alpha _{1}^{2}+{{b}^{-2}}-1 \right)}^{{1}/{2}\;}} \right]}^{2}}};
\end{equation}
\begin{equation}
\label{eq12}
D\left( {{\alpha }_{2}} \right)=\frac{4c{{\alpha }_{2}}{{\left( \alpha _{2}^{2}+{{b}^{2}}-1 \right)}^{{1}/{2}\;}}}{b{{\left[ {{\alpha }_{2}}+\left( c/b \right){{\left( \alpha _{2}^{2}+{{b}^{2}}-1 \right)}^{{1}/{2}\;}} \right]}^{2}}}.
\end{equation}
For a spherical Fermi surface

\begin{equation}
\label{eq13}
\begin{matrix}
  {{\left\langle \alpha  \right\rangle }_{{{v}_{z}}>0}}=1/2;\quad {{\left\langle \alpha D(\alpha ) \right\rangle }_{{{v}_{z}}>0}}=\int\limits_{0}^{1}{\alpha D(\alpha )}d\alpha ; \\
  R_{het}^{-1}=2R_{0}^{-1}\int\limits_{0}^{1}{\alpha D(\alpha )}d\alpha.  \\
\end{matrix}
\end{equation}
Since $R_{het}^{-1}$ does not depend on the number of the metal for which the averaging is carried out, we have
\begin{equation}
\label{eq14}
\begin{matrix}
2{{\left\langle {{\alpha }_{2}}D\left( {{\alpha }_{2}} \right) \right\rangle }_{{{v}_{z}}>0}}=2\int\limits_{\sqrt{1-{{b}^{2}}}}^{1}{{{\alpha }_{2}}D\left( {{\alpha }_{2}} \right)d{{\alpha }_{2}}}\equiv \\
\equiv2{{b}^{2}}{{\left\langle {{\alpha }_{1}}D({{\alpha }_{1}}) \right\rangle }_{{{v}_{z}}>0}}
\end{matrix}
\end{equation}

The integration for the second metal is performed from $\sqrt{1-{{b}^{2}}}$, because of the fact that for ${{\alpha }_{2}}< \sqrt{1-{{b}^{2}}}$ the quantity $D\left( {{\alpha }_{2}} \right)=0$, since total internal reflection of the electrons from the hetero boundary occurs.

Then using the equation \hyperref[eq3]{3} for the diameter of the homocontact in the form of a circular hole and equations \hyperref[eq13]{13}, \hyperref[eq14]{14} an expression for the diameter of the heterocontact follows:
\begin{equation}
\label{eq15}
\begin{matrix}
  {{d}_{het}}={{d}_{1}}{{\left[ 2{{\left\langle {{\alpha }_{1}}D({{\alpha }_{1}}) \right\rangle }_{{{v}_{z}}>0}} \right]}^{-1/2\;}}= \\ ={{d}_{2}}{{b}^{-1}}{{\left[ 2{{\left\langle {{\alpha }_{1}}D({{\alpha }_{1}}) \right\rangle }_{{{v}_{z}}>0}} \right]}^{-1/2\;}}
\end{matrix}
\end{equation}

Here $d_1$ is the diameter of the homocontact of a metal with a smaller value of the Fermi momentum, and $d_2$ is the diameter of the homocontact of a metal with a larger value of the Fermi momentum.

The use of diameters $d_1$ and $d_2$ is a convenient technique for calculating the heterocontact diameter ${d}_{het}$. It demonstrates that it is possible to calculate the homocontact diameter for any of the metals that compose the heterocontact and have a matching resistance, and then correct this diameter to the heterocontact diameter. Also, this technique clearly demonstrates that ${d}_{het}$ is always larger than $d_1$ or $d_2$ due to the reflection of part of the electron trajectories from the hetero-boundary, and the result of the calculation does not depend on the side of which metal the calculation was performed. 

\emph{This feature allows one to estimate its diameter by comparison with another heterocontact under certain conditions, even if the Fermi parameters of one of the metals are unknown. Suppose we have 3 different metals $A$, $B$ and $C$. For metals $A$ and $B$ the Fermi parameters are known, and for $C$ there is no information. Then for the heterocontact $A-B$, using the equation \hyperref[eq15]{15} it is easy to calculate its diameter. If for both of these heterocontacts by any experimental method (for example, by switching to a superconducting state one of the electrodes) to find the value of the barriers at the hetero-boundaries and these barriers are close to each other, we can assume that their diameters at close resistances will not differ much from each other. In the following, we will use this remark to estimate the diameter of the $NbSe_{2}-Cu$ heterocontact. Of course, such an estimate can only be qualitative, but it is better than nothing.}

Let us consider for illustration the $Ta-Cu$ heterocontact. In the free-electron approximation:

\begin{equation}
\label{eq16}
{{k}_{F}}={{\left( {3{{\pi }^{2}}z}/{\Omega }\; \right)}^{{1}/{3}\;}},
\end{equation}

where $z$ is the number of conductivity electrons per primitive cell, $\Omega$ is the volume of the primitive cell. For a VCC lattice, $\Omega ={{{a}^{3}}}/{2}$\, $a$ is the lattice constant. Using the free-electron approximation, the true wave functions are approximated by smooth pseudowave functions. The greatest differences are observed in the region of the core of the atom, which in simple metals is small and occupies about 10\% of the volume. In transport phenomena, in particular electrical conductivity, the free-electron approximation in such metals as copper, gold and silver "works" very well.

In the VA subgroup for $V$, $Nb$ and $Ta$ over the filled shell configuration of argon, krypton and xenon, respectively, there are 5 valence electrons per atom. Due to the small number of electrons filling the d-zones, the Fermi level crosses them, so the band structure of these metals near the Fermi surface is very complex. All metals of the subgroup are uncompensated with a total number of carriers of one hole per atom \cite{13}. Therefore, for tantalum we take $\emph{z}$=1 in the free-electron approximation.

Note that $\emph{z}$ is not always an integer. In \cite{13} there are values of $\emph{z}$ for a large number of transition metals, which can be used in estimates of this kind \cite{14}.

Given that $\emph{a}$=0.3296 nm, we find $k_{F}^{Ta}=1.183\cdot {{10}^{8}}\,c{{m}^{-1}}$. From the de Haase - van Alphen \cite{15} experiments, the ratio of the effective electron mass averaged over the Fermi surface to the free-electron value ${{{m}^{*}}}/{{{m}_{0}}=1.85}$ is determined for tantalum; then $v_{F}^{Ta}=0.74\cdot {{10}^{8}}cm/\sec$. For copper respectively: $v_{F}^{Cu}=1.57\cdot {{10}^{8}}cm/\sec$; $k_{F}^{Cu}=1.36\cdot {{10}^{8}}\,c{{m}^{-1}}$; then for copper and tantalum contact diameters we have respectively: 
$d_{Ta}=37.54\cdot R(\Omega)^{-1/2}[nm]$; $d_{Cu}=33.5\cdot R(\Omega)^{-1/2}[nm]$;

Then, as follows from the equations $b={p_{F}^{Ta}}/{p_{F}^{Cu}}={k_{F}^{Ta}}/{k_{F}^{Cu}}$=0.87; $c={v_{F}^{Ta}}/{v_{F}^{Cu}}$=0.544;

\begin{equation}
\label{eq17}
\begin{matrix}
%\begin{split}
{d}_{het}={{d}_{Ta}}{{\left[ 2{{\left\langle {{\alpha }_{1}}D({{\alpha }_{1}}) \right\rangle }_{{{v}_{z}}>0}} \right]}^{{-1}/{2}\;}}= \\ 
={{d}_{Cu}}{{b}^{-1}}{{\left[ 2{{\left\langle {{\alpha }_{1}}D({{\alpha }_{1}}) \right\rangle }_{{{v}_{z}}>0}} \right]}^{-1/2\;}}= \\
=70.2{{\left( R\left[ \Omega  \right] \right)}^{{-1}/{2}\;}}[nm],
%\end{split}
\end{matrix}
\end{equation}

The maximum angle of incidence of the electrons at the interface on the copper side is ${\theta }_{2}^{max}=60.46^\circ$.

Instead of the free-electron value $\rho l$ (Eq.\hyperref[eq2]{2}) we can use values obtained from experiments: $\rho l_{Ta}=0.59\cdot1{0}^{-11}\Omega\cdot c{m}^{2}$ \cite{16}; $\rho l_{Cu}=0.53\cdot1{0}^{-11}\Omega\cdot c{m}^{2}$ \cite{17}. For the contact diameters of copper and tantalum we have respectively: $d_{Ta}=31.65\cdot R(\Omega)^{-1/2}[nm]$; $d_{Cu}=30\cdot R(\Omega)^{-1/2}[nm]$; then from the Eq.\hyperref[eq2]{2} follows: $b={p_{F}^{Ta}}/{p_{F}^{Cu}}=({{\rho {l}_{Cu}/{\rho {l}_{Ta}}}})^{1/2}$=0.948; Fermi velocity ratio is the same: $c={v_{F}^{Ta}}/{v_{F}^{Cu}}$=0.544; ; then $d_{het}=59.2\cdot R(\Omega)^{-1/2}[nm]$ and the maximum angle of incidence of electrons at the interface on the copper side is ${\theta}_{2}^{max}=84^\circ$ .

In Figure {\hyperref[Fig1]{1} for the $Ta-Cu$ heterocontact, the angular dependence of the passage coefficient $D(\theta _{1})$ (equation \hyperref[eq11]{(11)}, converted to degrees) through the barrier at the hetero-boundary on the tantalum side in the free-electron approximation, as well as the associated tunneling parameter by the relation: $Z=\sqrt{(1/D)-1}$.

\begin{figure}[]
\includegraphics[width=8.5cm,angle=0]{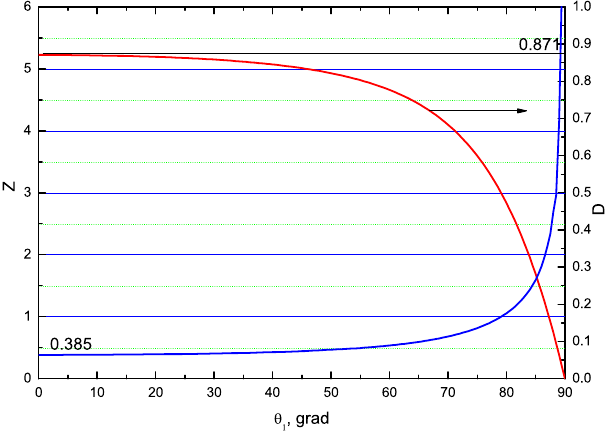}
\caption[]{Dependences of the tunneling parameter $Z$ and the transmittance coefficient $D$ on the deviation from the normal of the angle of incidence of the electrons on the heterogeneous boundary when calculated from the tantalum shore side in the free-electron approximation.}
\label{Fig1}
\end{figure}

Let us now consider what happens to the heterocontact when one of the banks transitions to the superconducting state.

In superconductor-normal-metal (hereafter $S-c-N$, here $c$ stands for constriction) point contacts with direct conduction, the current flowing is determined by a quantum process called Andreev reflection. In this process, the electron moving from the normal metal to the superconductor as it moves away from the heterogeneous interface at the coherence length is converted to a Cooper pair. Toward the electron, a hole from the opposite spin band passes into the normal metal. In the ideal case, in the absence of electron scattering at the boundary, at $T\rightarrow 0$ for voltages less than $\Delta/e$ the conductivity of the point contact doubles.

An intermediate, between tunneling and barrier-free, mode of current flow in point contacts is described by the Blonder-Tinkham-Klapwijk (BTK) model \cite{18}.

The following equation \hyperref[eq18]{(18)} (taken from \cite{19}, equation 5) is a modified version of the BTK equation in the two-gap approximation, which includes the possibility to account for finite carrier lifetime by introducing the $\Gamma$ broadening parameter. It is used to find the superconducting gaps $\Delta _{1}$ and $\Delta _{2}$, the broadening parameters $\Gamma_{1}$ and $\Gamma_{2}$, and the tunneling parameter $Z$. The dimensionless parameter $Z$ characterizes the value of the $\delta$-functional barrier at the boundary and can vary from 0 to infinity, in fact, at $Z\sim$10 we have a tunnel contact. The broadening parameter $\Gamma$ has the same dimensionality as the energy superconducting gap and leads to broadening and suppression of the intensity of the curves. The contribution to the total conductance from the $\Delta _{1}$ gap is given by the $K$ parameter and from the $\Delta _{2}$ gap by the $1-K$ parameter, respectively.
\begin{equation}
\label{eq18}
\frac{dV}{dI}=\frac{{S}_{F}}{\frac{dI}{dV}\left( \Delta _{1},\Gamma
_{1},Z\right) K+\frac{dI}{dV}\left( \Delta _{2},\Gamma
_{2},Z\right) \left( 1-K\right) }
\end{equation}

In the fitting process, using these parameters, the best match between the shape of the theoretical and experimental curve is achieved. Quantitatively, the fitting process consists in finding the minimum average standard deviation (the minimum RMS deviation) of points on the calculated curve from points on the experimental curve.

In this case, if the experimental curve of the differential resistivity of the superconducting state is normalized to the curve of the normal state, the scaling factor $S_F$ can be obtained from equation \hyperref[eq18]{(18)}. This is not a fitting parameter, but an indicator of how well the intensity (or amplitude) of the experimental curve matches the theoretical model. This factor is always present in the output computational data set when computing the average standard deviation, it is only necessary to ensure that it is directly output along with the output of the other computational parameters. If the amplitude of the experimental curve coincides with the theoretically expected one, then $S_F$=1; usually the amplitude of the experimental curve turns out to be smaller than the theoretical one, $S_F$<1, due to some reasons, e.g., not the whole volume is filled with superconductor, or a part of the superconductor has reduced superconducting characteristics, etc. However, in some cases it is possible to obtain $S_F$>1. This happens when the curves are strongly blurred (the parameter $\Gamma$ is comparable or larger than $\Delta$, and the average standard deviation varies small in a wide enough range of fitting parameters. In this case, sometimes its minimum value is achieved at $S_F$>1. Obviously, in this case one should choose such a set of parameters at which $S_F$<1.

The corresponding example can be seen in Figure 18 in \cite{20}. As can be seen from this figure, the average standard deviation $F(\Delta,\Gamma,Z)$ varies very small in the interval 0.6<$\Delta$<1.2 mV. As a rule, values with a scaling factor $S_F$>1 can be discarded. Also, it follows from this figure that if we consider only such sets of $Z$, $\Delta$, and $\Gamma$ for which $S_F=const$ (in Fig. 18 these are 1, 0.65, and 0.3), the uncertainty in finding $Z$, $\Delta$, and $\Gamma$ decreases dramatically. Therefore, usually, if there is a set of experimental curves whose blurring varies for whatever reason, first fit the curve with the minimum blurring and then, assuming that $S_F$>1 for the subsequent curves remains constant, find the superconducting parameters for the remaining curves. This generally works well for temperature measurements if no phase transition, such as a change in magnetic order, occurs in the range under investigation. This allows to reduce the error in calculations of temperature dependences of gaps in the region of high temperatures, where the experimental curves are strongly blurred. One should also understand that if there are two closely located superconducting gaps, or the gap varies along the Fermi surface, but not widened due to the finite lifetime of carriers, then the fit in the single-gap approximation with the parameter $\Gamma$ can be a good match in the shape of the theoretical and experimental curve, but the value of $S_F$ will be overestimated and may even exceed 1.

The BTK model and its modifications are one-dimensional, assuming that the charge carriers hit the boundary between the metals along a perpendicular trajectory. Nevertheless, it is perfectly suited, in particular, to find the $Z$ parameter in $Ta-Cu$ heterocontacts, given that it follows from Fig.\hyperref[Fig1]{1} that the appreciable growth of $Z$ (blue curve) begins at angles over $70^\circ$ from the perpendicular to the contact plane.

There is a more complex three-dimensional Zaitsev model in which the transparency coefficient $D$ can depend on the angle of incidence of the carriers at the interface \cite{21}. A review of \cite{22} in Section 2 also gives the Zaitsev equations, and Section 7 shows that applying the 3D model gives essentially the same result as the 1D model, except for the slightly different parameter $Z$ (see Figure 11, \cite{22}).

\section{Experimental Procedure}\label{sec3}

Point contacts were created between the massive electrodes. Single crystals of tantalum, copper, and $2H-NbSe_2$ were used as electrode materials. The criterion for the quality of the material used in point contact spectroscopy is the ratio of the resistivity at room temperature to the residual resistance at low temperature $\rho_{300}/\rho_{res}$. For a large number of metals and compounds there are known temperature-independent constants $\rho l$, where $l$ is the free path of carriers. Knowing these values, it is easy to estimate the impulse free path length at low temperature, which will be the estimate from above for the elastic electron path length through the point contact. For example, for our tantalum samples $\rho_{300}/\rho_{res}\sim20$, $\rho l=0.59\cdot {{10}^{-11}}\Omega \cdot c{{m}^{2}}$ \cite{16}, ${{\rho }_{273}}=12.6\cdot {{10}^{-6}}\Omega \cdot cm$\ \cite{23}, then the free path in the vicinity of the $Ta-Cu$ point contact cannot be greater than 90 nm.

To create ballistic point contacts it is necessary to use a technology that minimizes the formation of additional scattering centers in the surface layer of the material in the vicinity of the short circuit. As experience shows, it is necessary to completely exclude mechanical processing when making electrodes - cutting, grinding, etc.

Copper and tantalum electrodes were cut on an electrical discharge machine in the form of 10$\div$15 mm long bars and $1\times 1\times$ or $1.5\times 1.5\times m{{m}^{2}}$ cross sections. For the $NbSe_2$ experiments, the copper electrodes were cut in the shape of a pyramid with a base of $1\times 1\times$ or $1.5\times 1.5\times m{{m}^{2}}$ and a height of 4$\div$5 mm. The defective layer on the electrode surface was removed by chemical or electrochemical treatment in a mixture of concentrated acids.

Let us emphasize the importance of this operation - in addition to the removal of the defective layer, the properties of the oxide on the surface are very important. The contact area of the electrodes is many orders of magnitude larger than the point contact area, the supporting oxide ensures its mechanical and electrical stability. The thickness of the oxide should be optimal so that the contact is sufficiently mechanically stable and, at the same time, to minimize the introduction of additional scatterers when creating the short circuit. In addition, its electrical properties are very important - no leakage currents should flow through it, parallel to the current through the point-contact. It is also necessary that there are no intermediate shunt conductive layers between the insulating oxide and the metal. For some metals, this problem has not yet been solved.

For copper and tantalum no difficulties have arisen. For the (electro)chemical polishing of tantalum, the mixture consisted of $\text{HF}\ \text{:}\ \text{HN}{{\text{O}}_{\text{3}}}\text{:}\ \text{HCl}{{\text{O}}_{\text{4}}}$ taken in equal volume ratios, and for copper, from $\text{HN}{{\text{O}}_{\text{3}}}\ \text{:}\ {{\text{H}}_{\text{3}}}\text{P}{{\text{O}}_{\text{4}}}\ \text{:}\ \text{C}{{\text{H}}_{\text{3}}}\text{COOH}$ in a 2:1:1 volume ratio.

The electrodes were then washed in distilled water, dried, and mounted in a \cite{24} point contact device. Surface quality control after (electro)chemical treatment was performed using an optical microscope in oblique light. The working surface should be free of dirt and off color. The rounding radius of the pyramidal apex was $r \leq 0.1 mm$.

The $3\times 5\times m{{m}^{2}}$ electrode was cut with a blade from a $NbSe_2$ single crystal of about $\sim 0.1 mm$ thickness and bonded with silver paste to a wire holder. Immediately before measurements, the top layers were removed, ensuring that the copper counterelectrode touched the inner, perfect layers. Note that on the natural growth faces of the single crystal superconductivity is usually partially suppressed.

The device for creating point contacts allowed to smoothly change the pressure force between the electrodes and move them relative to each other \cite{24}. To ensure stability of the contacts, one of the electrodes is attached to a damper.

The $Ta-Cu$ contacts were created using the shift method \cite{10,25} in two steps. First, the electrodes were touched by the edges and then shifted relative to each other. The resistance of the resulting contacts was continuously monitored. Contacts with a resistance of several hundred ohms to several kilohms were selected for the next stage. By regulating the strength of the electrodes pressed against each other, such contacts were obtained quite often. Then with the help of the decade resistor connected in series with the voltage source and the found point contact we began to increase the current in steps. Resistance of the point contact also decreased in steps.

The breakdown voltage at the contact was $500\pm200$ mV. When the desired resistance interval was reached, the contact was held under the final current for several minutes. Resistances of good quality point contacts obtained by this method ranged from 30-40 to 200-250 $\Omega$, the quality criterion being the EPI spectra. The highest parameters of spectra showed point contacts with resistance of 50-80 $\Omega$ \cite{10}. The point contacts obtained by this method were of much higher quality than those obtained by the standard shift method and had better mechanical and electrical stability. 

The $Cu-NbSe_{2}$ point contacts were created using the standard shift method \cite{25} - the top of the copper pyramid was pressed against the $NbSe_{2}$ surface with a small force and then shifted parallel. Varying, if necessary, the pressing force, we obtained a point contact for subsequent measurements.

\section{Experimental Results}\label{sec4}
\subsection{Superconducting gap and nonequilibrium feature in $Ta-Cu$ point contacts}\label{subsec1}

For the experimental estimation of the barrier value in the heterocontact due to the mismatch of the Fermi parameters, the point contacts should be ballistic and have no additional scatterers in the contraction plane. The BTK equation \cite{18} and its modification \cite{19} refer to the ballistic mode of electron flight through the point contact. As shown in \cite{26}, in the diffusion mode the first derivative of the $I-V$ curves with parameter $Z=0$ practically coincides in form with that in the ballistic mode with tunneling parameter $Z=0.55$. The corresponding illustration can be seen in the overview \cite{21} in Fig. 9.

One can distinguish the diffusion contact from the ballistic contact by the appearance of the second derivative of the $I-V$ curves. The decrease in the elastic electron scattering length is due to an increase in the number of scatterers, i.e., an increase in the concentration of impurities and lattice defects, which leads to distortion of the crystal lattice of the metal. And since nonequilibrium phonons reflect the vibrational structure of the material in the vicinity of its generation, as the elastic relaxation length decreases, there is a broadening of the EPI peaks in the spectra, the suppression of high-energy phonon features up to their complete disappearance and to the growth of the background. In \cite{27} the effect of Nb point contact contamination on the EPI spectra was considered.

\begin{figure}[]
\includegraphics[width=8.5cm,angle=0]{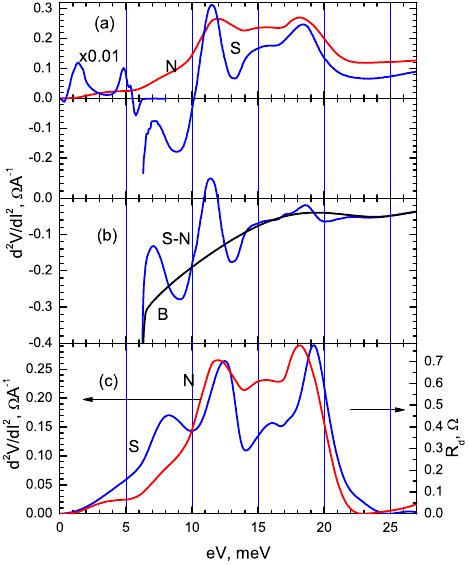}
\caption[]{(a) Second derivative of the $I-V$ curve of the $Ta-Cu$ point contact in the normal ($N$) and superconducting ($S$) states. The initial area of the $S$-spectrum is reduced in intensity by a factor of 100. $H$=0, $T_{N}$=4.3 K, $T_{S}$=1.7 K, $R_{0}^{N}$=73$\Omega$. (b) $S-N$ is the difference between the second derivatives of the $I-V$ curve, B is the background curve. (c) $N$ is the second derivative of the $I-V$ curve of $Ta-Cu$ point contact in the normal state after the background correction; $S$ is the differential resistance proportional to the EPI function obtained from the superconducting addition to the spectrum after subtracting the background curve ($S-N-B$).}
\label{Fig2}
\end{figure}

A more complicated case is the identification of the scatterers on the heterogeneous boundary. The influence of the translucent boundary wall on the appearance of the second derivative of the $I-V$ curve (T-model) is considered in \cite{28}. It shows that the intensity of the phonon peaks in the spectra is inversely proportional to the transparency coefficient. At the same time, the intensity of the two-phonon processes decreases much slower. Thus, the relative intensity of the two-phonon processes on the second derivatives in the normal state can be used to judge the presence of such a boundary.

Note that the low intensity of the EPI spectra in the absence of their broadening and low background level is not an unambiguous sign of the T-model and may be due to multi-contact (small-diameter contacts included in parallel have a lower spectrum intensity than a single point contact with the same resistance), or a strong deviation of the short circuit shape from the circular hole (for example, a long crack in the backing oxide).

Thus, the simplest test, a kind of passport characterizing the mode of electron passage through the point contact, is the form of the second derivative of the $I-V$ curve in the normal state.

\begin{figure*}[t!]
\centering
\includegraphics[width=1\linewidth]{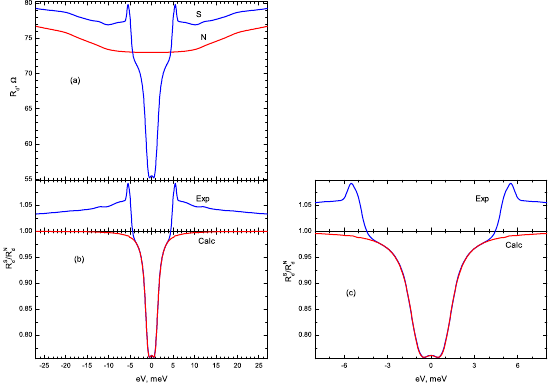}
\caption{
(a) - differential resistances of the $Ta-Cu$ heterocontact shown in Fig.\hyperref[Fig2]{2}, in normal ($N$) and superconducting ($S$) states. (b) - differential resistance after normalization (Exp), and the calculated curve (Calc). (c) - the curves shown in (b) on a larger scale. T=1.7 K, $\Delta$=1.04 meV, $Z$=0.307, $\Gamma$=0.38 meV, $S_{F}$=0.99555}
\label{Fig3}
\end{figure*}

Fig.\hyperref[Fig2]{2(a),(b)} shows the second derivatives of the $I-V$ curves of the $Ta-Cu$ point contact in the normal and superconducting states, as well as the difference curve and superconducting background curve, and Fig. \hyperref[Fig2]{2(c)} are the curves proportional to the EPI function obtained from these spectra. The procedure for correcting the background and restoring the EPI function from the superconducting additive to the spectrum is described in detail in \cite{27}. The large intensity of high-energy phonon peaks and pronounced van Hove features, unequivocally testify to the ballistic flight of electrons through the point contact and unperturbed tantalum crystal structure in the volume, on the order of the coherence length, where the formation of phonon nonlinearity in the superconducting state \cite{5} occurs.

Let us now turn to the initial region of the second derivative of the superconducting state point contact $I-V$ curves (Fig. \hyperref[Fig2]{2(a)}). Along with the nonlinearity due to the energy gap $\Delta$ in the spectrum of quasiparticle excitations, there is a feature on the curve due to the jump-like transition of the superconductor to a nonequilibrium state when the critical concentration of nonequilibrium quasiparticles in the near-contact region \cite{1,2} is reached.

For different contacts, the position of such features depends on their resistance, temperature, and/or external magnetic field. During the transition to the superconducting state, the initiation of such features occurs near the characteristic phonon energies (low-frequency phonon mode, the first or second phonon peak, depending on the contact resistance). As the temperature decreases, their intensity increases, and they shift to lower energies.

\begin{figure*}[t!]
\centering
\includegraphics[width=1\linewidth]{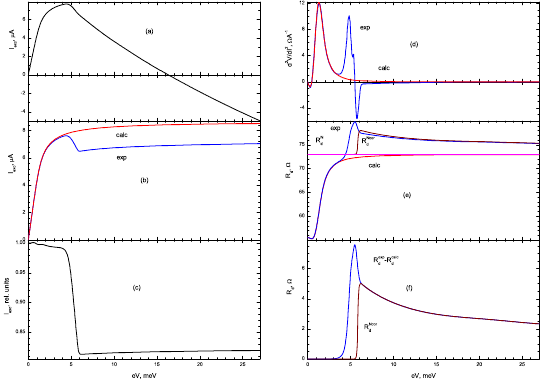}
\caption{
(a) Excess current calculated using experimental differential resistance curves, for superconducting and normal states, shown in Fig.\hyperref[Fig3]{3(a)}. (b) The same, for the curves shown in (e). (c) The relative value of the excess current from the bias at the contact. (d) are the second derivatives of the $I-V$ curve obtained after subtracting the nonlinearities due to the EPI. The $exp$ curve in the initial section coincides with the curve $S$ in Fig.\hyperref[Fig2]{2(a)}, and at $eV$>6 meV coincides with the background curve $B$ in Fig.\hyperref[Fig2]{2(b)}. The $calc$ curve is obtained by differentiating the calculated curve in Fig.\hyperref[Fig3]{3(b)} and scaled accordingly. (e) $exp$ and $calc$ are the differential resistances corresponding to the second derivatives in panel (d), $R_{d}^{N}$=73 $\Omega$ is the normal state differential resistance for curve $calc$, $R_{d}^{Ncor}$ -- corrected normal state differential resistance for curve $exp$, details in text. (f) The difference resistance differential and the corrected normal curve $R_{d}^{Ncor}$.}
\label{Fig4}
\end{figure*}

At a fixed temperature, the position of the features on the energy axis is proportional to $R^{1/2}$, which corresponds to the constancy of the critical power ${{{P}_{c}}=V_{c}^{2}}/{R\simeq\text{const}}\;$ ($\simeq 0.4\mu \text{W}$ at 2 K).

The effect of the magnetic field is similar to that of temperature - the feature is blurred, its intensity decreases, and it shifts to the region of higher energies. The corresponding temperature and magnetic-field dependences are shown in Figs. 8-10 in \cite{2}.

Since reaching the critical concentration of quasiparticles above the gap depends on the ratio of the rate of their generation, determined by the power, and the recombination (escape) rate, which increases with temperature and magnetic field, this explains the similar temperature and magnetofield dependence of the position of features on the energy axis.

Fig.\hyperref[Fig3]{3} shows the differential contact resistances in the normal and superconducting states (panel (a)), as well as the normalized $Exp=R_{d}^{S}/R_{d}^{N}$ curve and the calculated $Calc$ curve (panel (b)). Panel (c) shows the curves from panel (b) on a larger scale. As follows from the parameters of the calculated $Calc$ curve shown in the figure, there is a good agreement in the value of the obtained tunneling parameter ($Z$=0.307) with the estimate of the same at the perpendicular electron falling on the interface ($Z$=0.385, fig.\hyperref[Fig1]{1})). The discrepancy $\sim20\%$ is apparently due to the rough estimation of the Fermi velocity ratio of the contacting metals. Hence, we can assume that the other estimates (e.g., for the diameter of the heterocontact) are also quite adequate. Also, based on the proximity of $S_F$ to 1, the superconducting properties of the contact are consistent with the theoretical model.

The shape of the nonequilibrium feature corresponds to a jump-like decrease in excess current and is accompanied by an increase in differential resistance. As is accepted in experimental work \cite{29}, by excess current here we mean the voltage-dependent difference between the $I-V$ curves in the $S$ and $N$ states.

If to calculate the value of excess current from bias on the contact use experimental curves of differential resistance in the normal $N$ and superconducting $S$ states (Fig.\hyperref[Fig3]{3(a)}), the excess current becomes negative (Fig.\hyperref[Fig4]{4(a)}) at voltage over 16 mV. This does not make physical sense and reflects the fact that the superconducting state $I-V$ curve has a larger slope and crosses the normal state $I-V$ curve. In order to correctly estimate the dependence of the excess current on the bias, it is necessary to take into account the change in this slope in the vicinity of the nonequilibrium singularity. For this purpose, let's find the differential contact resistance dependence on the bias in the superconducting state without taking into account the EPI.

In Fig.\hyperref[Fig4]{4(d)} the second derivatives of the $I-V$ curve are shown, in which there are no spectral components. The $exp$ curve in the initial section coincides with the curve $S$ in Fig. \hyperref[Fig2]{2(a)}, and when the voltage exceeds 6 mV it coincides with the background curve $B$ in Fig.\hyperref[Fig2]{2(b)}. The $calc$ curve is obtained by differentiating the calculated curve in Fig.\hyperref[Fig3]{3(b)} and scaled accordingly.

Panel \hyperref[Fig4]{4(e)} shows the differential resistances corresponding to these curves, as well as the differential normal state resistances for the calculated $calc$ curve, which is a horizontal line at 73$\Omega$, and the corrected differential normal state resistivity curve for the experimental $exp$ curve. It consists of three parts. The initial segment coincides with the straight line , the second segment represents the difference between the differential resistances of the experiment and the calculation at a bias greater than 6 mV (panel \hyperref[Fig4]{f}). The stepped segment conjugates the two parts $R_{d}^{Ncor}$ . Such an unusual, at first sight, choice of the shape of this curve is connected with the necessity to exclude the influence of the differential resistance jump ($I-V$ curve kink) when finding the excess current.

The differential resistance difference from the contact bias shown in panel \hyperref[Fig4]{f} shows a maximum around 5.5 mV, a value of 7.58 $\Omega$ or $\sim$10\% of the contact resistance in the normal state at zero bias. This value is an order of magnitude greater than the spectral component proportional to the EPI function (Fig.\hyperref[Fig5]{5}).

\begin{figure}[]
\includegraphics[width=8.5cm,angle=0]{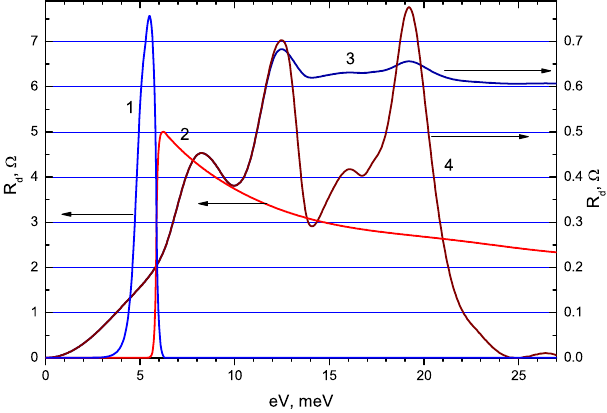}
\caption[]{Comparison of the relative magnitude of additional nonlinearities in the superconducting state associated with nonequilibrium processes in point contacts and with EPI: 1 - jump of differential resistance of the point contact, corresponding to the reduction of excess current during the transition of the superconducting region near the contact to the nonequilibrium state; 2 - additional differential resistance arising as a result of imbalance of occupancy of hole and electronic branches of quasiparticle excitation spectrum, appearance of reverse current and related additional voltage; 3 - change of differential resistance of point contact due to scattering of andreef electrons on nonequilibrium phonons; 4 - the same, after background correction. }
\label{Fig5}
\end{figure}

Fig.\hyperref[Fig4]{4(b)} shows the dependences of excess current values on the contact bias $calc$ and $exp$, calculated from the differential resistance curves shown in panel \hyperref[Fig4]{4(e)}, and panel \hyperref[Fig4]{4(c)} shows the relative value of excess current drop when the near-contact region enters the nonequilibrium state. As follows from the figure, the excess current suppression is less than 20\%.

Qualitative explanation of the increase in the differential resistance of the point contact during the transition to a nonequilibrium state is based on the appearance of reverse current and the associated additional voltage that increases the contact resistance due to the imbalance of the occupancy of hole and electronic branches of the excitation spectrum of quasiparticles \cite{1,2}.

Through the $N-S$ boundary, quasiparticles with maximum energy $eV\gg\Delta$ are injected into the superconductor, which populate the electron-like or hole-like branches of the excitation spectrum, depending on the polarity of the applied voltage. The excitations relax relatively quickly, emitting phonons and accumulating in a layer on the order of $\Delta$ above the ceiling of the energy gap. Further relaxation of the residual population unbalance of excess quasiparticles occurs rather slowly, over a time of the order of ${{\tau }_{0}}\sim {{\tau }_{ep}}(\Delta )$ \cite{7}, during which the excitation manages to diffuse deep into the superconductor to a distance $\lambda_{Q} \sim {{\left( {{l}_{i}}{{l}_{\Delta }} \right)}^{{1}/{2}\;}}$, where ${{l}_{\Delta }}={{v}_{F}}{{\tau }_{0}}$.

Since the potential difference falls at a distance of the order of $d$ from the contact plane, similar to what happens in tunneling $S-I-N$ contacts, in the near-contact region the chemical potential of quasiparticles is not equal to the chemical potential of pairs. Here there is an excess charge of quasiparticles and the associated reverse current (or added voltage), which increases the contact resistance the greater the charge value. Factors that reduce the magnitude of the unbalance (inelastic scattering on phonons, superconducting current, etc.) reduce the reverse current and contact resistance.

The shape of the differential resistance curve at voltages higher than the critical voltage is determined by the dependence of the relaxation rate on $eV$. As the voltage increases, the injection increases, but at the same time the relaxation rate also increases. The residual unbalance of the population imbalance of the excitation spectrum branch of excess quasiparticles due to the increasing frequency of electron-phonon scattering decreases, which leads to a decrease in the differential resistance of the point contact with respect to the differential resistance in the normal state (see Fig. \hyperref[Fig3]{3(b)}). 

\subsection{Superconducting gap and nonequilibrium feature in $2H-NbSe_{2}-Cu$ point contacts}\label{subsec2}

Part of the results presented in this subsection was previously published in \cite{30}, dedicated to the consideration of spatially inhomogeneous discrete states of superconductors. Here this material is significantly expanded, the emphasis is placed on the importance of taking into account the non-equilibrium processes in the observed effects.

$NbSe_{2}$ is a layered easily split superconductor with a very high degree of anisotropy. It is formed by three-layer "sandwiches": selenium layer - niobium layer - selenium layer. In each layer, atoms form a tightly packed triangular lattice. The lattice parameters $a\approx$0.345 nm; $c\approx$1.254 nm; the lattice period along the $c$ axis contains two monolayers \cite{31}. The strong anisotropy here is due to the weak van der Waals interaction between the selenium layers closest to each other, located in different structural sandwiches.

 In the normal state at room temperature $\rho_{\parallel}\sim$2$\cdot$10$^{-4}\Omega\cdot$cm; $\rho_{\bot}\sim$10$^{-3}\Omega\cdot$cm \cite{32} which is two orders of magnitude greater than the resistivity of typical metals. The coherence length for the unperturbed material is: $\xi_{\parallel}\simeq$7.8 nm, $\xi_{\bot}\simeq$2.6 nm, i.e., perpendicular to the layers is almost the same as the lattice period.
 
The equation \hyperref[eq18]{18} assumes a ballistic mode of electron flight through the point contact. While this requirement can be met with respect to the impulse and energy electron path lengths within the framework of the materials and technology used to create point contacts, this is not possible with respect to the coherence lengths due to natural reasons, especially for the $c$ direction. Nevertheless, due to the lack of an alternative, we will use the equation \hyperref[eq18]{18} when finding the values of the energy gaps).

\begin{figure}[]
\includegraphics[width=8.5cm,angle=0]{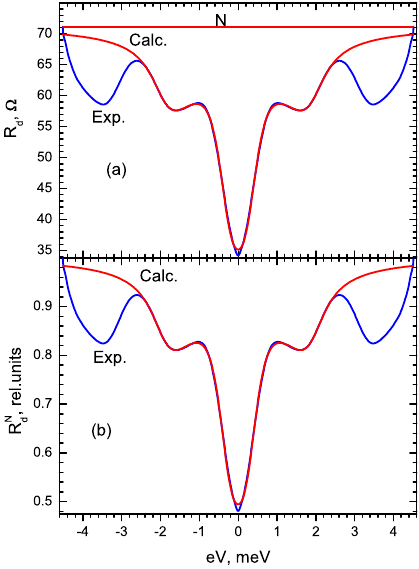}
\caption[]{(a) - differential resistance of point contact $NbSe_{2}-Cu$. Blue curve - experiment, red calculation, $N\sim$71 $\Omega$.
(b) - curves after normalization. $\Delta_{1}$=0.245 meV; $\Delta_{2}$=1.915 meV; $\Gamma_{1}$=0.0232 meV; $\Gamma_{2}$=0.1 meV; $Z$=0.346; contribution from gap $\Delta_{1}$, K=0. 835; gap contribution $\Delta_{2}$, (1-K)=0.165, (see Eq.\hyperref[eq15]{15}); $S_{F}$=1.702; $2\Delta_{1}/kT_{C}$=0.8; $2\Delta_{2}/kT_{C}$=6.24.}
\label{Fig6}
\end{figure}

The differential resistance of the $NbSe_{2}-Cu$ point contact in the superconducting state ($Exp$ curve) as well as the theoretical curve ($Calc$) calculated within the framework of the modified BTK model are shown in Fig.\hyperref[Fig6]{6(a)}. Unfortunately, we do not have the normal-state curve. Nevertheless, we managed to find this curve for further evaluations using a simple procedure. Using parameters of the curve $Calc$, we calculated exactly the same curve, but already normalized to the normal state with some scaling factor $S_{F}$. After that we divided the original calculated curve by the last one. By the method of successive approximations, varying $S_{F}$, we achieved that as a result of such division a horizontal segment of a straight line is obtained. Its position on the ordinate axis corresponds to the value of resistance of the point contact in the normal state.

The energy gap values obtained as a result of the fitting correlate well with those obtained, for example, with tunnel contacts, see, e.g., \cite{33}. In spite of the fact that the values of superconducting energy gap $\Delta_1$ and $\Delta_2$ differ practically 8 times, and one would assume an appreciable difference of Fermi electron parameters in different zones, nevertheless it was possible to obtain an excellent agreement on the form of calculated and experimental curve using the same for both gaps tunneling parameter $Z$ in the fitting process.

The deviation in shape from the calculated curve at displacements over 2.5 mV is apparently due to the presence at 3$\div$6 mV of a group of phonons associated with charge density waves (CDWs) \cite{34}. Static distortions of the lattice caused by CDWs result in a superstructure with a period approximately triple that of the original lattice. The occurrence of the superlattice leads to the aforementioned low-energy CDWs phonons.

In \cite{35} it is shown that for superconductors with strong EPI for contacts with direct conductivity, the elastic component of the current leads to additional nonlinearity associated with the dependence of the superconducting gap on the bias at the contact and due to the electron-phonon renormalization of the energy spectrum of the superconductor. It is also shown there that elastic processes lead to the appearance of maxima of differential conductivity in the region of characteristic phonon energies on the first derivative of the excess current, which is observed in our experiment.

Note that previously we observed a similar manifestation of elastic scattering processes in lead and indium \cite{4,36} point contacts. It is important to emphasize that along with elastic scattering processes, in superconducting contacts with direct conductivity also coexist inelastic phonon scattering processes on Andreev electrons, which leads to a reduction of the excess current. That is, in this case, phonon features manifest themselves in the form of differential resistance maxima of the excess current. Thus, these contributions are directed oppositely to each other and can mutually weaken. Moreover, it is difficult to estimate in advance which contribution will be predominant, much depends on external conditions.

As for the use of the same tunneling parameter $Z$ for both purposes in two-gap superconductors (see Eq.\hyperref[eq18]{(18)}), we used this approach to study the gap structure in nickel borocarbide compounds \cite{19,22,37,38} and obtained an excellent agreement between the experimental and calculated characteristics of the point contacts studied.

Let's pay attention to the fact that the tunneling parameter of our point contact ($Z$=0.346) is very close to the theoretical estimate for the point contact $Ta-Cu$ with perpendicular electrons falling on the interface ($Z$=0.385, Fig.\hyperref[Fig1]{1}). The point contact resistances are also very close to each other ($R_{N}$=73$ \Omega$ for $Ta-Cu$ and $R_{N}$=71$ \Omega$ for $NbSe_{2}$).

As noted in the theoretical section of the paper, for two different heterocontacts in which one of the metals is the same and the tunneling parameters are close, the diameters of these contacts do not differ much from each other at close resistances. Therefore, we can assume that the diameters of $Ta-Cu$ and $NbSe_2-Cu$ contacts are very close to each other, the estimate being $d\sim$8.5 nm. We remind that this is a qualitative estimate, which may differ markedly from the true size.

Since the coherence length perpendicular to the layers is approximately the same as the lattice period in the same direction, and the lattice period contains 2 monolayers, it follows from this estimate that at least four $NbSe_{2}$ monolayers adjacent to the hole fall into the current concentration region. To summarize, the ballistic condition with respect to the coherence length is clearly violated, while at the same time the elastic and inelastic relaxation lengths are noticeably larger than the contact size. 

The applicability of modified BTK theory to these kinds of contacts turned out to be quite satisfactory except for the intensity of the spectra: the scale factor $S_{F}$ was 1.7 times the theoretical expectation, which manifested itself in a doubling of the differential resistance at the transition to the normal state (Fig.\hyperref[Fig6]{6(a)}) $R_{0}\approx$35 $\Omega$; $R_{N}\approx$70 $\Omega$).

\begin{figure}[]
\includegraphics[width=8.5cm,angle=0]{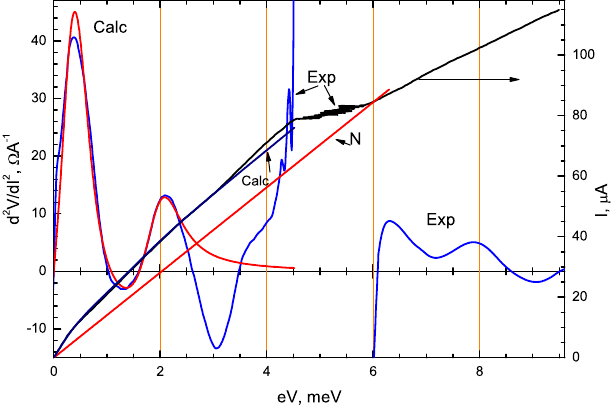}
\caption[]{$I-V$ curve and its second derivative for the $NbSe_{2}-Cu$ point contact, $T$=1.7 K, $H$=0. $Exp$ -- experimental data, $Calc$ -- calculated curve, $N$ -- normal state approximation (see Fig.\hyperref[Fig6]{6}). }
\label{Fig7}
\end{figure}

\begin{figure}[]
\includegraphics[width=8.5cm,angle=0]{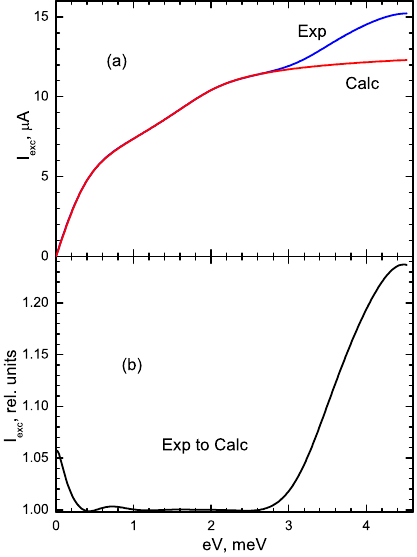}
\caption[]{(a) - excess current for the curves shown in Fig.\hyperref[Fig7]{7}.
(b) is the relative magnitude of the excess current from the bias.}
\label{Fig8}
\end{figure}

Fig.\hyperref[Fig7]{7} shows the second derivative and the $I-V$ curve of the same contact in a wider energy range. As can be seen from the figure, at voltages above 4.5 mV the transition of the $I-V$ curve to a new branch with a large differential resistance is observed. The transition mechanism here is similar to that in $Ta-Cu$ point contacts. Electrons with excess energy $eV$, scattering on low-energy CDWs-phonons, lose energy and accumulate above the gap. In tantalum, the concentration growth of nonequilibrium quasiparticles occurs in a large volume with a size on the order of the coherence length ($\xi_{0}\sim$90 nm), which promotes a smooth phase transition to the suppressed gap state.

In the case of $NbSe_{2}$, however, the transition to the nonequilibrium state occurs in the layer adjacent to the contact and located in the current concentration region. In this case, the smallest fluctuations in the current strength caused by external inductions lead to fluctuations in the concentration of nonequilibrium quasiparticles, which manifests itself in the corresponding form of the $I-V$ curve. After reaching the critical concentration and transition of two monolayers into the nonequilibrium state with a suppressed gap, the $I-V$ curve moves to a branch with a large differential resistance (see also Fig.\hyperref[Fig9]{9(b)}, $N\sim$71 $\Omega$ -- linear approximation of the normal state differential resistance, $R_{d}\sim$120$ \Omega$ -- quasilinear approximation of the new branch), similar to what took place for the $Ta-Cu$ contact.

\begin{figure}[]
\includegraphics[width=8.5cm,angle=0]{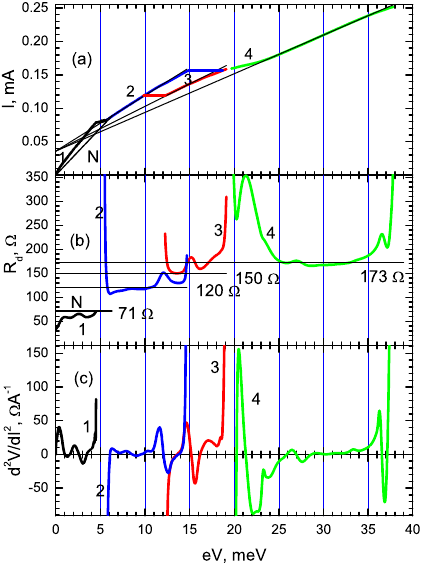}
\caption[]{(a) $I-V$ curve of the $NbSe_{2}-Cu$ point contact, $T$=1.7 K. The step plots are marked with numbers 1-4. Tangents to them are drawn by thin lines. $N$ is the normal state line.
(b) Differential resistances for the plots of the panel $I-V$ curve (a). Thin lines correspond to the tangents for these sections.
(c) Second derivatives for the corresponding sections.}
\label{Fig9}
\end{figure}

In Fig.\hyperref[Fig8]{8}, panel (a) shows the experimental and calculated dependences of the excess current before the transition of the superconductor into a nonequilibrium state, in panel (b) - the experimental value normalized to the calculated value. As follows from the figure, the increase of the superconducting gap due to elastic processes of electron-phonon reformation of the superconductor energy spectrum, leads to an increase in the excess current by approximately 24\% compared to the calculation.

Fig.\hyperref[Fig9]{9} shows the $I-V$ curves, first and second derivatives of the same point contact in the whole bias range. In panel (a) tangents are drawn to the sections of the $I-V$ curves, designated by numbers 2, 3 and 4. As can be seen from the figure, on these quasi-linear sections of the $I-V$ curve the differential resistance changes in jumps, resembling the features caused by the formation of phase slip centers in a thin superconducting filament. The differential impedance of the tangent sections of the $I-V$ curve 2, 3, and 4 (panel (a)) is 120, 150, and 173 $\Omega$ (see panel (b)); the incremental differential impedance is 30 and 23 $\Omega$.

The reason for the appearance of such a stepped structure of the $I-V$ curve is the layered structure of $2H-NbS{e}_{2}$. While strong covalent chemical bonds are present within each layer, neighboring layers are held together by the much weaker van der Waals interaction. As already noted, the coherence length perpendicular to the layers practically coincides with the lattice period containing 2 monolayers. And since the conversion of Andreev electrons to Cooper pairs occurs at a distance of the order of the coherence length, the maximum value of the superconducting current is reached at the boundary of the lattice period. In fact, due to the weak coupling between the layers, when the value of the critical current is exceeded, the weak coupling begins to generate a flux of normal quasiparticles.

Since the energy range of quasiparticles with energy $0<\epsilon<eV$ is significantly larger than the contact size, the inelastic relaxation in the second pair of layers united by a common coherence length causes the non-equilibrium quasiparticles to accumulate above the gap, but their concentration is still less than critical to pass to the non-equilibrium state. Therefore, the transition of the Josephson coupling between the first and second pairs of layers into a resistive state sharply increases the flux of nonequilibrium quasiparticles into the second pair and switches it into the nonequilibrium state with the suppressed gap.

Note the large hysteresis loop between curves 2 and 3; the figure shows the maximum loop obtained. During multiple recordings, branch-to-branch breaks could occur under the action of the leads at other points as well. 
The hysteresis loop arises because before the switching we had an asymmetric Josephson transition, in which in the first pair the gap was suppressed and in the second pair it had an equilibrium state. After switching the second pair, we had a symmetric Josephson transition with suppressed gap, and the reversal of the $I-V$ curve goes on a branch with a large differential up to the second pair transition again to the state with the equilibrium gap. Note, the transition from branch to branch for the maximum hysteresis loop occurs around the maximums of differential resistance of point contacts on the first derivative of the $I-V$ curve (panel b). These maxima correspond to the phonon state density maxima. Near these phonon peaks, there is a faster change in the concentration of nonequilibrium quasiparticles above the gap due to scattering of quasiparticles with maximum energy $eV$ on nonequilibrium phonons. Switching from branch to branch at other points of this hysteresis loop is due to random inductions.

Thus, the successive transition of the three pairs of layers adjacent to the contact into a nonequilibrium state with a suppressed gap is accompanied by a change in the quasilinear differential resistance from 71 $\Omega$ (normal state approximation) to, respectively, 120, 150, and 171 $\Omega$ and an incremental decrease of 49, 30, and 23 $\Omega$. The number of pairs of $2H-NbS{e}_{2}$ layers that have transitioned to the nonequilibrium state gives us an independent estimate of the contact diameter. Three pairs in the current concentration region give an estimate for a diameter of $d\sim$15 nm. The initial qualitative estimate of the diameter is only about half as good as this independent estimate, demonstrating the validity of such estimates.

\section{Discussion}\label{sec5}

For the phase transition of a superconductor to the nonequilibrium state with a suppressed gap, it is necessary and sufficient to increase the concentration of nonequilibrium quasiparticles above the critical one above the gap in a layer of order $\Delta$ above the ceiling of the energy gap. To achieve the critical concentration, double tunneling contacts are often used. In such structures, one of the contacts is a low-resistance tunneling junction that creates a nonequilibrium superconducting state (generator) in the middle film. The second contact is higher impedance to introduce a minimum of perturbations into the middle film and serves to obtain information about this state (detector) \cite{39}. Due to the geometry of the experiment, varying the flux of nonequilibrium quasiparticles with the desired energy is quite simple.

If we consider the geometry of the experiment with respect to $NbSe_2-Cu$ point contacts, it is much closer to tunnel injection experiments than three-dimensional $Ta-Cu$ contacts. The role of the generator here is played by a copper electrode, van der Waals bonds between the $NbSe_2$ layers prevent the current from flowing and promote the successive transition of the layers to a nonequilibrium state when the critical concentration of quasiparticles is reached in them. 

Once again, let us return to the validity of the application of the BTK equation in finding the parameter $Z$ and the correspondence of this parameter to the barrier arising from the mismatch of the Fermi parameters of the contacting metals. It is important to realize that such a comparison can only be qualitative, but the comparison with experiment, in particular, for the $Ta-Cu$ contact gives a discrepancy of $\sim20\%$, which allows us to conclude that for isotropic metals with close values of $\rho l$ this estimate can be quantitative if the Fermi velocity ratios are correctly found. At the same time, one should not forget that all this is valid only under the condition of ballistics and the absence of additional scatterers at the hetero-boundary. Therefore, for each such determination of the parameter $Z$ it is necessary to obtain the EPI spectrum of the contact in the normal state and make sure that there are distinct phonon peaks, etc., which was discussed in section 4.

It seems unexpected, but even for such a strongly anisotropic superconductor as $NbSe_2$ the estimation of the diameter of the $NbSe_2-Cu$ contact by comparing it with the $Ta-Cu$ contact having a close value of the tunneling parameter $Z$ gave a value only twice different from the estimation of the diameter by the number of pairs of layers transitioning to the nonequilibrium state. Taking into account that the transition to the nonequilibrium state of the layers is possible not only in the region of current concentration, but also beyond its limits due to a sufficiently long energy relaxation length of electrons, as it takes place for tantalum contacts, it can be assumed that the initial estimate of the diameter probably gives a smaller error. At least, these estimates complement each other and allow us to obtain more objective information.

Finally, we once again draw attention to the fact that nonequilibrium features were never observed in dirty point contacts, the perfection of the superconductor crystal lattice plays a very important role. For example, non-equilibrium features were absent in the spectra in dirty tantalum-based point contacts. A similar observation applies to niobium-based point contacts. The presence of a clear step structure of the HTSC $I-V$ curve associated with the discrete character of the electric field penetration into the region of the point contact constriction was also observed in samples with a high degree of crystalline order in the \cite{30,40} constriction.

\section{Conclusion}\label{sec6}

\begin{enumerate}
\item It was found that after the transition of the superconductor region to a nonequilibrium state, this state turns out to be stable to changes in the injection power (the excess current and, consequently, the energy gap, change very insignificantly in a wide range of biases.

\item It was found that the transition of the superconductor region to the nonequilibrium state with a reduced gap is possible only for an unperturbed superconductor with a perfect lattice.
    
\item It is shown that the increase in the differential resistance of the point contact during the transition to a nonequilibrium state occurs due to the appearance of unbalanced occupancy of the hole and electronic branches of the quasiparticle excitation spectrum, which leads to the appearance of reverse current and the additional voltage associated with it.

\item It is found that the use of the modified BTK equations for pure superconducting point contacts with a coherence length smaller than the diameter, leads to overestimated values of the amplitude (or intensity) of the gaps.

\item The possibility of estimating the value of the normal resistance of a point contact using its superconducting characteristics is shown.
\end{enumerate}

The work was supported by the National Academy of Sciences of Ukraine within the F19-5 project.


\begin{thebibliography}{}
\makeatletter
\renewcommand{\@biblabel}[1]{#1.}
\makeatother

\bibitem{1}\href{https://fnt.ilt.kharkiv.ua/index.php/fnt/article/view/f12-0552r/2119}{I.K.~Yanson, L.F.~Rybal’chenko, N.L.~Bobrov, and V.V.~Fisun, Fiz. Nizk. Temp. \textbf{12}, 552 (1986)} [Sov. J. Low Temp. Phys. \textbf{12}, 313 (1986)], \href{https://doi.org/10.48550/arXiv.1512.00684}{arXiv.1512.00684}
\bibitem{2}\href{https://fnt.ilt.kharkiv.ua/index.php/fnt/article/view/f13-1123r/2420}{I.K.~Yanson, N.L.~Bobrov, L.F.~Rybal’chenko, V.V.~Fisun, Fiz. Nizk. Temp. \textbf{13}, 1123 (1987)} [Sov. J. Low Temp. Phys. \textbf{13}, 635 (1987)] \href{https://doi.org/10.48550/arXiv.1512.03917}{arXiv.1512.03917}
\bibitem{3}\href{https://fnt.ilt.kharkiv.ua/index.php/fnt/article/view/f09-0985r/1617}{V.A.~Khlus, Fiz. Nizk. Temp. \textbf{9}, 985, (1983)} [Sov. J. Low Temp. Phys. \textbf{9}, 510 (1983)]
\bibitem{4}\href{https://DOI.org/10.3367/UFNe.2019.11.038693}{N.L.~Bobrov, UFNe, \textbf{63}(11), 1072, (2020)} [\href{https://DOI.org/10.3367/UFNr.2019.11.038693}{UFNr, \textbf{190}(11), 1143, (2020)}], \href{https://doi.org/10.48550/arXiv.2109.00806}{arXiv:2109.00806v1 [cond-mat.supr-con] }
\bibitem{5}\href{https://doi.org/10.1063/1.5097356}{N.L.~Bobrov, Low Temperature Physics \textbf{45}, 482 (2019)}; \href{https://fnt.ilt.kharkiv.ua/index.php/fnt/article/view/f45-0562r/8046}{Fiz. Nyzk. Temp.  \textbf{45}, 562-575 (2019)}.  \href{https://doi.org/10.48550/arXiv.1906.04380}{arXiv.1906.04380} 
\bibitem{6}\href{https://doi.org/10.1103/PhysRevLett.39.229}{R.C.~Dynes, V.~Narayanamurti, and J.P.~Garno Phys. Rev. Let. \textbf{39}, N4, 229 (1977)} 
\bibitem{7}\href{https://doi.org/10.1103/PhysRevB.14.4854}{S.B.~Kaplan, C.C.~Chi, D.N.~Langenberg, et al., Phys. Rev. B \textbf{14}, 4854 (1976).}
\bibitem{8}\href{http://www.jetp.ras.ru/cgi-bin/dn/e_021_03_0655.pdf}{Yu.V.~Sharvin JETP, \textbf{21}, No3, 655 (1965)}
\bibitem{9} Kulik~I.O., Omel'yanchuk~A.N, Shekhter~R.I. \emph{Sov. J. Low Temp. Phys.} \textbf{3} 740 (1977);\href{https://fnt.ilt.kharkiv.ua/index.php/fnt/article/view/f03-1543r/566}{\emph{Fiz. Niz. Temp.} \textbf{3} 1543 (1977)}
\bibitem{10}\href{http://fnt.ilt.kharkiv.ua/index.php/fnt/article/view/f13-0611r/2323}{N.L.~Bobrov, L.F.~Rybal'chenko, V.V.~Fisun, I.K.~Yanson, Fiz. Nizk. Temp., \textbf{13}, 611 (1987)}; Sov. J. Low Temp. Phys., 12, 344 (1987); \href{https://doi.org/10.48550/arXiv.1512.01800}{arXiv.1512.01800} 
\bibitem{11}\href{http://fnt.ilt.kharkiv.ua/index.php/fnt/article/view/f09-0046r/1486} {R.I.~Shekhter and I.O.~Kulik, Fiz. Nizk. Temp. \textbf{9}, 46 (1983)} [Sov. J. Low Temp. Phys. 9, 22 (1983)].
\bibitem{12}\href{https://en.wikipedia.org/wiki/Fermi_energy}{WikipediA. Fermi energy}
\bibitem{13}\href{https://doi.org/10.1002/pssb.2220540150}{G.~Bambakidis, Phys. Status Solidi (b) \textbf{54}, K57 (1972)}.
\bibitem{14}\href{https://doi.org/10.1063/1.4963330}{N.L.~Bobrov, L.F.~Rybal'chenko, V.V.~Fisun, V.V.~Khotkevich, Low Temperature Physics \textbf{42}, 811 (2016)}; [\href{https://fnt.ilt.kharkiv.ua/index.php/fnt/article/view/f42-1035r/7541}{Fiz. Nyzk. Temp., \textbf{42}, 1935 (2016)}]; \href{https://doi.org/10.48550/arXiv.1612.03396}{arXiv:1612.03396 [cond-mat.mes-hall]}.
\bibitem{15}\href{https://doi.org/10.1103/PhysRevB.1.366}{M.H.~Halloran, J.H.~Condon, J.E.~Graebner, J.E.~Kunzier, F.S.L.~Hsu, Phys. Rev. B \textbf{1}, 366, (1970).} 
\bibitem{16}\href{https://doi.org/10.1007/BF00654497}{V.V.~Ryazanov, V.V.~Schmidt and L.A.~Ermolaeva, J. Low Temp.Phys. \textbf{45}, No. 5/6, 507 (1981)}. 
\bibitem{17}\href{https://doi.org/10.1103/PhysRevB.31.8244}{M.J.G.~Lee, J.~Caro, O.G.~Croot, R.~Griessen. Phys. Rev. B \textbf{31}, No. 12, 8244 (1985).} 
\bibitem{18}\href{https://doi.org/10.1103/PhysRevB.25.4515}{G.E.~Blonder, M.~Tinkham, and T.M.~Klapwjik, Phys. Rev. B \textbf{25}, 4515 (1982).}
\bibitem{19}\href{https://doi.org/10.1103/PhysRevB.71.014512}{N.L.~Bobrov, S.I.~Beloborod’ko, L.V.~Tyutrina, I.K.~Yanson, D.G.~Naugle, and K.D.D.~Rathnayaka, Phys. Rev. B \textbf{71}, 014512 (2005)}; \href{https://doi.org/10.48550/arXiv.cond-mat/0412325}{arXiv:cond-mat/0412325 [cond-mat.suprcon]}
\bibitem{20}\href{https://fnt.ilt.kharkiv.ua/index.php/fnt/article/view/f36-1228e/6561}{N.L.~Bobrov, V.N.~Chernobay, Yu.G.~Naidyuk, L.V.~Tyutrina, I.K.~Yanson, D.G.~Naugle, K.D.D.~Rathnayaka, Fiz. Niz. Temp., \textbf{36}, N10-11, p.1228-1243, (2010)} [\href{https://doi.org/10.1063/1.3521569}{Low Temperature Physics \textbf{36}, 990-1003 (2010)}; \href{https://doi.org/10.48550/arXiv.1006.5933}{arXiv:1006.5933 [cond-mat.supr-con]}
\bibitem{21}\href{http://www.jetp.ras.ru/cgi-bin/dn/e_059_05_1015.pdf}{A.V.~Zaitsev, Sov. Phys. JETP \textbf{59}, 1015 (1984)} [Zh. Eksp. Teor. Fiz. 86, 1742 (1984)]
\bibitem{22}\href{https://doi.org/10.1063/1.5030447}{Yu.G.~ Naidyuk, K.~ Gloos, Low Temperature Physics \textbf{44}, 257–268 (2018)} [\href{http://fnt.ilt.kharkiv.ua/index.php/fnt/article/view/f44-0343e/7827}{Fiz. Nyzk. Temp., \textbf{44}, 343-356 (2018)}.]
\bibitem{23}V.E.~Startsev, "Local singularities in the Fermi surfaces and electronic transport phenomena in transition metals" Author's Abstract of Doctoral Dissertation in Physical-Mathematical Sciences, Sverdlovsk (1983)
\bibitem{24}\href{https://patents.su/5-1631626-ustrojjstvo-dlya-polucheniya-okhlazhdaemogo-tochechnogo-kontakta-mezhdu-metallicheskimi-ehlektrodami.html}  {N.L.~Bobrov, L.F.~Rybal’chenko, A.V.~Khotkevich, and I.K.~Yanson, USSR Patent No. \textbf{1631626}, Byull. Izobret., No. 8 (1991).} 
\bibitem{25}\href{https://patents.su/2-834803-sposob-polucheniya-prizhimnykh-mikro-kohtaktob-mezhdu-metallicheskimiehlektrodami.html}{P.N.~Chubov, A.I.~Akimenko, and I.K.~Yanson, USSR Patent No. \textbf{834803}, Byull. Izobret., No. 20 (1981), p. 232.} 
\bibitem{26}\href{https://doi.org/10.1063/1.1357127}{I.I.~Mazin, A.A.~Golubov, and B.~Nadgorny, J. Appl. Phys. \textbf{89}, 7576 (2001).} 
\bibitem{27}\href{https://doi.org/10.1134/S1063776121060108}{N.L.~Bobrov, J. Exp. Theor. Phys. \textbf{133}, 59–70, (2021).} \href{https://doi.org/10.48550/arXiv.2109.01344}{arXiv.2109.01344} 
\bibitem{28}\href{https://doi.org/10.1016/0038-1098(78)90916-X}{A.P.~van~Gelder, Solid State Comm., \textbf{25}, No 12, 1097 (1978)} 
\bibitem{29}\href{https://doi.org/10.1103/PhysRevB.27.112}{G.E. Blonder and M. Tinkham, Phys. Rev. B \textbf{27}, 112 (1983).}
\bibitem{30}I.K. Yanson, L.F. Rybal'chenko, V.V. Fisun, N.L. Bobrov, M.A. Obolenskii, M.B. Kosmyna, V.P. Seminozhenko, Sov. J. Low Temp. Phys., \textbf{14}, 639 (1988) \href{http://fnt.ilt.kharkiv.ua/index.php/fnt/article/view/f14-1157r/2649}{Fiz. Nizk. Temp., \textbf{14}, 1157 (1988)}; \href{https://doi.org/10.48550/arXiv.1512.06416}{arXiv.1512.06416} 
\bibitem{31}\href{https://doi.org/10.1063/1.3423025}{I.A.~Gospodarev, V.V.~Eremenko, K.V.~Kravchenko, V.A.~Sirenko, E.S.~Syrkin, and S.B.~Feodos’ev, Low Temp. Phys. \textbf{36}, 344 (2010)} \href{https://fnt.ilt.kharkiv.ua/index.php/fnt/article/view/f36-0436r/6470}{Fiz. Nizk. Temp., \textbf{36}, 436 (2010)}
\bibitem{32}\href{https://doi.org/10.1016/S0022-3697(71)80400-6}{J.~Edwards, R.F.~Frindt, J. Phys. Chem. Solids \textbf{32} 2217 (1971)}
\bibitem{33}\href{https://doi.org/10.1038/s41467-018-03000-w}{T.~Dvir, F.~Massee, L.~Attias, M.~Khodas, M.~Aprili, C.H.L.~Quay, H.~Steinberg, Nat Commun \textbf{9}, 598 (2018).}
\bibitem{34}\href{https://fnt.ilt.kharkiv.ua/index.php/fnt/article/view/f11-0925r/1969}{N.L.~Bobrov, L.F.~Rybal’chenko, M.A.~Obolenskii, and V.V.~Fisun, Fiz. Nizk. Temp., \textbf{11}, 897 (1985)}; (Sov. J. Low Temp. Phys., \textbf{11}, 510 (1985)) DOI: \href{https://doi.org/10.48550/arXiv.1603.02598}{arXiv.1603.02598} 
\bibitem{35}Beloborod’ko S.I., Omel’yanchuk A.N.,  Sov. J. Low Temp. Phys. \textbf{14} 630 (1988); \href{https://fnt.ilt.kharkiv.ua/index.php/fnt/article/view/f14-0322r/2506}{Fiz. Niz. Temp. \textbf{14} 1142 (1988)}
\bibitem{36}\href{https://doi.org/10.1063/1.4869565}{N.L.~Bobrov, A.V.~Khotkevich, G.V.~Kamarchuk, P.N.~Chubov, Low Temp. Phys. \textbf{40}, 215 (2014)}; \href{https://fnt.ilt.kharkiv.ua/index.php/fnt/article/view/f40-0280r/7114}{Fiz. Niz. Temp. \textbf{40}, 280, (2014)}  \href{https://doi.org/10.48550/arXiv.1405.6869}{arXiv.1405.6869} 
\bibitem{37}\href{https://doi.org/10.1063/1.2199452}{N.L.~Bobrov, S.I.~Beloborod’ko, L.V.~Tyutrina, V.N.~Chernobay, I.K.~Yanson, D.G.~Naugle, K.D.D.~Rathnayaka, Low Temperature Physics \textbf{32}, 489 (2006)}; \href{https://fnt.ilt.kharkiv.ua/index.php/fnt/article/view/f32-0641e/5801}{Fiz. Niz. Temp. \textbf{32}, 641, (2006)} \href{https://doi.org/10.48550/arXiv.cond-mat/0511373}{arXiv.cond-mat/0511373}
\bibitem{38}\href{https://doi.org/10.1209/0295-5075/83/37003}{N.L.~Bobrov, V.N.~Chernobay, Yu.G.~Naidyuk, L.V.~Tyutrina, D.G.~Naugle, K.D.D.~Rathnayaka, S.L.~Bud'ko, P.C.~Canfield, I.K.~Yanson, EPL \textbf{83} 37003 (2008)} \href{https://doi.org/10.48550/arXiv.0806.1456}{arXiv.0806.1456}
\bibitem{39}\href{https://doi.org/10.1063/1.3699014}{E.M.~Rudenko Low Temp. Phys. \textbf{38}, 353 (2012)} \href{https://fnt.ilt.kharkiv.ua/index.php/fnt/article/view/f38-0451r/6804}{Fiz. Nizk. Temp., \textbf{38}, 451 (2012)}
\bibitem{40} L.F.~Rybal'chenko, V.V.~Fisun, N.L.~Bobrov, M.B.~Kosmyna, A.I.~Moshkov, V.P.~Seminozhenko, I.K.~Yanson, Sov. J. Low Temp. Phys., \textbf{15}, 54 (1989); \href{http://fnt.ilt.kharkiv.ua/index.php/fnt/article/view/f15-0095r/2697}{Fiz. Nizk. Temp., \textbf{15}, 95 (1989)}; \href{https://doi.org/10.48550/arXiv.1701.09124}{arXiv.1701.09124} 
\end{thebibliography}
\end{document}